
\input phyzzx.tex

\overfullrule=0pt
\singlespace

\def\cl{\centerline}
\def\etal{{\it et al.\ }}
\def\bs{\bigskip}

\Pubnum{\hfil SCIPP 92/51}
\date{\hfil October 1992}
\titlepage

\vskip 1cm

\cl{\bf PARTICLE ASTROPHYSICS AFTER COBE}
\smallskip
\cl{\bf BLOIS92 SUMMARY TALK}

\bs
\cl{JOEL R. PRIMACK}
\cl{Santa Cruz Institute for Particle Physics}
\cl{University of California, Santa Cruz, CA 95064}

\bs\bs
\cl{\bf Abstract}

The IV Rencontres de Blois, on Particle Astrophysics, held
at the Ch\^ateau de Blois, June 15-20, 1992, was a meeting
well-timed for a reconsideration of the issues in particle
astrophysics in the light of the COBE discovery of the
cosmic microwave background (CMB) fluctuations.  This is a
summary of what I thought were the most interesting things
discussed at Blois: (1) The near-success of Cold Dark Matter
(CDM) in predicting the COBE fluctuation amplitude, which
favors the hypothesis that structure formed in the universe
through gravitational collapse. (2) The indications that
$\Omega\approx1$ and that the power spectrum has a little
more power on supercluster and larger scales than CDM.
These are suggested by the IRAS and CfA redshift surveys and
POTENT galaxy peculiar velocity analysis, and also by the
COBE data. (3) The consequent demise of CDM and the rise of
hybrid schemes such as Cold+Hot Dark Matter (C+HDM). (4) The
possible implications for neutrino masses and mixings, and
for cosmology, of the recent results on solar neutrinos.
(5) The first discovery of TeV $\gamma$ rays from an
extragalactic source, which was announced at Blois.

I also summarize here a number of the exciting ongoing and
planned experiments and observations discussed at Blois:
CERN experiments on $\nu_\mu \nu_\tau$ oscillations, which
may be sufficiently sensitive to detect the $\nu_\tau$ if
its mass lies in the cosmologically interesting mass range
1-10$^2$ eV; dark matter searches, including the French and
Berkeley-Livermore-Mt. Stromlo MACHO search and the searches
for WIMPs and axions; and the construction of ambitious
laser interferometer gravity wave detectors such as LIGO in
the U.S. and VIRGO in France and Italy.  The commitment of
funds for VIRGO was announced at Blois by the French
Minister of Research and Space, Hubert Curien.

\endpage
\singlespace
\bs
\noindent {\bf I. Introduction}

The goal of Particle Astrophysics is to construct a
fundamental theory of the material universe---i.e., to
explain elegantly and economically observations from the
smallest physical scale to the entire cosmological horizon.
Of course, science can never tell us which theories are
``true''; at best it can only tell us which are false.
Paradoxically, the theories that most closely approach truth
are those whose limits are {\it known}, like Newtonian
Mechanics.  We know where Newtonian Mechanics is valid
because we know precisely where and how it fails, since it
is enveloped on all sides by more accurate theories: Quantum
Mechanics for small sizes, Special Relativity for high
speeds, General Relativity for large size or large
gravitational potential $\phi \sim m/r$.\ref{C.W. Misner
1977, in {\it Cosmology, History, and Theology} W. Yourgrau
\& A.D. Breck, eds. (Plenum Press), p. 75.} It has even been
possible to combine some of these theories, as in QED.  But
we do not know where or how these theories in turn fail.  So
our goal is to construct enveloping theories for them.

In particular, in particle physics our goal is to construct
an enveloping theory for the 3-2-1 ``Standard Model,'' based
on the SU(3)$_c\times$[SU(2)$\times$U(1)]$_{ew}$ gauge group
for three generations of quarks and leptons, with all three
neutrinos massless.  A century ago, there were only a few
``clouds on the horizon'' portending the storms that
destroyed classical physics.  Perhaps the main cloud now on
the horizon of the particle physics Standard Model is the
hint of neutrino mass from the solar neutrino data, which I
will summarise in \S IV.

In cosmology, we do not yet even have a fundamental theory.
Cosmology today is like physics before Newtonian Mechanics
or geology before Plate Tectonics.  We only have bits and
pieces of the story.  Perhaps standard Big Bang
Nucleosynthesis is such a piece.  Almost certainly General
Relativity is.  Cold Dark Matter (CDM) was an educated guess
regarding such a fundamental theory.  But, like the
original SU(5) Grand Unified Theory in particle physics, CDM
was apparently too simple to be true, as I will summarize in
\S II-III. In constructing a fundamental theory of
cosmology, it now appears that the data requires a hybrid
theory containing elements of at least two simpler theories,
such as Cold + Hot Dark Matter (C+HDM; see \S III). The
resulting theory may thus be a little like the 3-2-1
standard model of particle physics, which is of course also
a hybrid theory. Another possibly useful analogy is
elliptical planetary orbits.  For millennia, until Kepler
and Newton, astronomical prejudice favored circles; now we
know that only an unusual accident of planetary formation
would give truly circular orbits (i.e., very small
ellipticity).

\noindent {\bf II. Cold Dark Matter}

In saying that the data do not favor the original CDM
theory, I do not mean to imply that there is any evidence
against all or most of the dark matter being of the ``cold''
variety, such as weakly interacting massive particles
(WIMPs) or axions.  I propose to use capital letters CDM (Cold
Dark Matter) to refer to the ``standard CDM'' theory based
not only on the assumption that the dark matter is cold, but
also on the assumptions that structure in the universe grew
gravitationally from Gaussian adiabatic fluctuations with a
Zel'dovich spectrum in a universe of critical density
($\Omega=1$).  These latter assumptions are of course just
what the simplest versions of inflation imply.\ref{Recent
reviews are A. Linde 1990, {\it Particle Physics and
Inflationary Cosmology} (Harwood); K. Olive 1990,
Phys. Rep. 190, 307.}

With CDM, the primordial Zel'dovich $|\delta_k|^2 \propto
k^n$ spectrum with $n=1$ is preserved on large scales but
tilted toward $n \rightarrow -3$ on short scales because
matter fluctuations that enter the horizon in the
radiation-dominated universe grow only
logarithmically.\ref{J.R. Primack \& G.R. Blumenthal 1984,
in {\it Clusters and Groups of Galaxies} F. Mar\-di\-ros\-sian,
G.~Giuricin \& M.~Mezzetti, eds. (D.~Reidel, Dordrecht)
p.~435; reprinted in {\it Particle Physics and Cosmology:
Dark Matter}, M. Srednicki, ed. (North-Holland, 1990), p.
90.}  The division between these regimes occurs at the
transition between radiation and matter domination, which
corresponds to a length scale of about 13
$(\Omega_0 h^2)^{-1}$ Mpc, or a mass scale of about $3.2
\times 10^{14} (\Omega_0 h^2)^{-2} \, M_\odot$.  This is
where the CDM fluctuation spectrum has a ``knee;'' for
lengths or masses larger than this, the amplitude of the
fluctuations starts to fall off rapidly, approaching the
primordial Zel'dovich spectrum on large length scales.

Standard CDM\ref{ G.R. Blumenthal, S. Faber, J.R. Primack,
\& M. Rees 1984, Nature 311, 517.} with biased galaxy
formation\ref{M. Davis, G. Efstathiou, C. Frenk \& S.D.M.
White 1985, Ap. J. 292, 371.} gives an excellent account of
structure formation from galaxy to cluster scales.  But as
Juskiewicz summarizes in these proceedings, all the
available evidence on large scale structure---from the
galaxy streaming velocities, APM\ref{S.J. Maddox \etal 1990,
M.N.R.A.S. 242, 43P.} and COSMOS\ref{C.A. Collins, R.C.
Nichol \& S.L Lumsden 1992, M.N.R.A.S. 254, 295.}
measurements of the galaxy angular correlation function
$w_g(\theta)$, IRAS and CfA redshift surveys (see also de
Lapparent, these proceedings), radio galaxy and rich
clusters data,\ref{J.A. Peacock \& M.J. West 1992,
MN.R.A.S., in press; S. Olivier, J.R. Primack, G.
Blumenthal \& A. Dekel 1993, Ap. J., in
press.} and now COBE---is pretty consistent.  And this evidence
suggests that a little more power is required on length
scales of $\sim10^2$ Mpc and beyond than that in the CDM
fluctuation spectrum---at least, if the visible matter is
related to the underlying mass distribution in a simple
way.\ref{For a recent review, see M. Davis, G.
Efstathiou, C.S. Frenk, and S.D.M. White 1992, Nature 356,
489.} CDM could perhaps be consistent with the data if
galaxy formation, like the weather, is such a complicated
process that it can only be described by a rather arbitrary
biasing prescription.\ref{E.g., A. Babul \& S.D.M. White
1991, M.N.R.A.S. 253, 31P.}  But as an originator of CDM, I
have always felt that one of its most attractive features is
its highly predictive character. So I consider such CDM
models to be non-standard, and not obviously more attractive
than other CDM variants such as $\Omega=0.2$ CDM or C+HDM,
for example.

To keep the situation in perspective, it is important to
note that CDM does not fail by very much.  COBE sees
$10^\circ$ fluctuations with rms amplitude over the whole
sky of $\Delta T/T \approx 10^{-5}$.  The amplitude
predicted by standard CDM is $10^{-5}/b$, where the biasing
factor $b$ is as usual the inverse of the rms mass
fluctuation in a sphere of radius 8 $h^{-1}$ Mpc.\ref{In a
sphere of this radius, the rms fluctuation in the number of
optically bright galaxies is unity. As usual, we use the
reduced Hubble parameter $h\equiv H_0/[100$ km s$^{-1}$
Mpc$^{-1}]$.} Thus with $b=1$, CDM agrees with COBE, and
also incidentally with much of the large scale data.
However, almost all nonlinear CDM calculations agree that
$b=1$ CDM predicts galaxy velocities on small scales that
are too high, while $b\approx2.5$ does much better in this
regard.  Thus the problem with CDM is only about a factor of
two or three.  But the COBE and large scale galaxy
distribution data are now so good that this sort of fudge is
unacceptable!  However, this near-agreement certainly does
suggest that some---perhaps most---of the basic assumptions
of CDM may be right.  In particular, it suggests that
structure grew in the universe by gravitational collapse
rather than, for example, because of energy input from giant
explosions: Matter fell, it wasn't pushed!

\noindent {\bf III. Hybrid Models for Large Scale Structure}

Perhaps the simplest variant of CDM that remains viable has
$\Omega\approx0.2$ with $h \approx 1$ and a cosmological
constant $\lambda \equiv \Lambda/3H_0^2 = 1-\Omega$ for
consistency with inflation and with CMB constraints.  This
model has more large scale power than standard CDM mainly
because matter domination occurs later with $\Omega$ lower,
so the ``knee'' in the power spectrum is moved to larger
scales.  This model is claimed to be consistent with the
galaxy angular correlation function $w_g(\theta)$,\ref{G.
Efstathiou, W.J. Sutherland \& S.J. Maddox 1990, Nature 348,
705} with the observed rich cluster correlation function
$\xi_c(r)$\ref{See also J.R. Primack in these proceedings,
and J. Holtzman \& J.R. Primack 1993, Ap. J., in press.} and
mass function,\ref{P. Lilje 1992, Ap. J. Lett. 386, L33; N.
Bahcall \& R. Cen 1992, Ap. J., in press.} and with power
spectra from clusters,\ref{R. Scaramella 1992, Ap. J. Lett.
390, L57.} the CfA slices (de Lapparent, these proceedings),
and the Southern Sky redshift survey.\ref{C. Park, J.R. Gott
\& L.N. da Costa 1992, Ap.J. Lett. 392, L51.}  There is a
possible problem in this model simultaneously fitting the
large-scale peculiar velocities, which require small linear
bias $b<1$, and COBE, which requires larger $b$.\ref{G.
Efstathiou, J.R. Bond \& S.D.M. White 1992, M.N.R.A.S. 258,
1P.}

There are, moreover, several indications that $\Omega \approx
1$, for example CMB dipole vs. QDOT/IRAS data, comparison of
IRAS density and galaxy peculiar velocity data,
reconstructing Gaussian initial conditions from the POTENT
analysis of galaxy peculiar velocity data, and void outflow
(see the talks by Dekel and Yahil in these proceedings).
While this evidence that $\Omega=1$ is still not compelling,
and the arguments for a large Hubble parameter\ref{J.P.
Huchra 1992, Science 256, 321; S. van den Bergh 1992,
Science 258, 421.} and an old universe do point toward
smaller $\Omega$, I personally am persuaded that it is
likely that $\Omega=1$.

The question arises whether {\it any} $\Omega=1$ model with
a physically motivated smooth spectrum of adiabatic Gaussian
fluctuations can account for all the data now available,
including the COBE CMB fluctuations (corresponding to scales
of 3000--300 h$^{-1}$ Mpc), large scale structure data
(300--10 h$^{-1}$ Mpc scales: galaxy angular correlations
$w_g(\theta)$, the cluster correlation function $\xi_c(r)$,
and galaxy streaming velocities, etc.), and smaller scale
structure data (10 h$^{-1}$ Mpc--10 h$^{-1}$ kpc: galaxy
formation, correlations, and velocities)?

One variant of standard CDM that has received much attention
recently\ref{E.g., R. Cen, N.Y. Gnedin, L.A. Kofman \& J.P
Ostriker 1992, Ap. J. Lett., in press.} keeps all the usual
assumptions except the Zel'dovich primordial spectrum
$|\delta_k|^2\propto k^n$ with $n=1$, substituting instead
``tilted'' spectra with $n\approx0.5-0.7$ that arise from
more or less complicated inflationary models.  Such models
have the virtue of being very well specified, with $n$ being
the only additional parameter beyond those of standard CDM.
However, the latest and most detailed studies\ref{F.C. Adams
\etal 1992, Fermilab preprint; A.R. Liddle \& D.H. Lyth
1992, Sussex preprint.} conclude that ``tilted'' CDM is
marginal at best.  For example, for $n<0.6$, sufficiently
small to account for the observed large scale structure,
there is probably too little early galaxy formation.  Of
course, it is possible to get much more general non-
Zel'dovich primordial fluctuation spectra from
inflation,\ref{For a review, see e.g. J.R. Primack 1991, in
{\it Proc. IUPAP Conf. Primordial Nucleosynthesis and
Evolution of Early Universe}, K. Sato, ed. (Kluwer), p.
193.} but these ``designer spectra'' are neither well
motivated nor well specified.

I will use the phrase Cold + Hot Dark Matter (C+HDM) to
refer to a model with $\Omega=1$ having roughly half as much
hot (light neutrino) dark matter as cold dark matter.  These
proportions of hot and cold dark matter are required to fit
the large-scale structure data (as I discuss further in my
contributed paper in these proceedings). C+HDM is physically
at least as well motivated as tilted CDM or any other
variant of CDM that we know.  Moreover it is well specified
and has only one additional parameter beyond those of
standard CDM: the neutrino mass $m(\nu_\tau)$, or
equivalently
$$\Omega_\nu=[m(\nu_\tau)/23\,{\rm eV}]h_{50}^{-2}\ ,
\eqn\omeganueq$$
where
$h_{50}(=2h)$ is the Hubble parameter $H_0$ in units of 50
km s$^{-1}$ Mpc$^{-1}$.\ref{With $\Omega=1$, the age of the
universe $t_0=13.04 h_{50}^{-1}$ Gy, so avoiding conflict
with Globular Cluster and other age estimates requires
$h_{50}\lsim 1$.} The required value of
$m(\nu_\tau)$, about 7 eV for $h_{50}\approx1$, is
consistent with the value implied by the currently available
solar neutrino data plus the old ``seesaw'' models of
neutrino masses, as I will discuss in \S IV below. The
neutrinos provide an unclustered dark matter component on
small scales, which could help explain why dynamical
estimates give $\Omega<1$ on small scales.  The
out-of-equilibrium relativistic Fermi-Dirac statistics of
the neutrinos\ref{Once the neutrinos decouple, their momenta
just redshift; see e.g., Weinberg 1972, {\it Gravitation and
Cosmology} (Wiley), p. 535.} enhances this effect.

The main objection to C+HDM in principle is the apparent
unliklihood of having two different dark matter components
each making comparable contributions to the mass density.
Although one of the earliest C+HDM papers\ref{K. Shafi \& F.
Stecker 1984, Phys. Rev. Lett. 53, 1292.} proposed a
particle physics model to account for this, I am unaware of
any such model that is attractive. However, the entire
particle physics Standard Model begs for further
explanation, so it should not disturb us to contemplate one
more feature that, if valid, would call for a more
fundamental justification.

Basic properties of mixed dark matter models were worked out
some time ago;\ref{L.Z. Fang, S.X. Li \& S.P. Xiang 1984,
Astr. Astrophys. 140, 77; R. Valdarnini \& S. Bonometto
1985, Astr. Astrophys. 146, 235; S. Achilli, F. Occhionero
\& R. Scaramella 1985, Ap. J. 299, 577.} and the fact that
C+HDM with $\Omega_{cdm}\approx0.6$ and $\Omega_\nu
\approx0.3$ is a promising model for large scale structure
was established by several previous linear
calculations.\ref{See my contributed talk in these
proceedings.}  The C+HDM power spectrum\ref{J. Holtzman
1989, Ap. J. Supp. 71, 1.} fits the data better than any
other model yet proposed.\ref{A.N. Taylor \& M.
Rowan-Robinson 1992, Nature 359, 396.} A simplified
nonlinear calculation in a 14~Mpc box has been done with the
initial neutrino fluctuations set equal to zero.\ref{M.
Davis, F.J. Summers \& D. Schlegel 1992, Nature 359, 393.}
My colleagues and I have just done the first detailed
nonlinear calculations for C+HDM, with proper initial
conditions, sufficiently many hot particles to sample
velocity space adequately, and a careful analysis of dark
matter and galaxy correlations and velocities with
comparisons to the available data.\ref{A. Klypin, J.
Holtzman, J.R. Primack \& E. Reg\H{o}s 1992, UCSC preprint.}
We find that C+HDM normalized with linear bias factor
$b=1.5$ is consistent both with the COBE data and with the
observed galaxy correlations. The number density of
galaxy-mass halos is only a little smaller than for CDM at
zero redshift but increasingly smaller at redshift $z>2$,
but the numbers of cluster-mass halos are slightly larger.
We also find that on galaxy scales the neutrino velocities
and flatter power spectrum in C+HDM result in galaxy
pairwise velocities that are in good agreement with the
data, and about 30\% smaller than in CDM with the same
biasing factor.  On scales of several tens of Mpc, the C+HDM
streaming velocities are considerably larger than CDM.  As a
result, the ``cosmic Mach number''\ref{J.P. Ostriker and Y.
Suto 1990, Ap. J. 348, 378; Y. Suto, R. Cen, and J.P.
Ostriker 1992, Ap. J. 395, 1.} in C+HDM is about a factor of
two larger than in CDM, and probably in better agreement
with observations.

Thus C+HDM looks promising as a model of structure
formation.  The presence of a hot component requires the
introduction of a {\it single} additional parameter beyond
standard CDM --- $m(\nu_\tau)$ or equivalently $\Omega_\nu$
--- and allows this model to fit essentially {\it all} the
available cosmological data remarkably well---except the
latest upper limit on $\sim 1^\circ$ CBM fluctuations from
the Santa Barbara South Pole experiment.\Ref\rGaier{ T. Gaier \etal
1992, Ap. J. Lett. 398, L1.}  It has been claimed that no
Gaussian model can simultaneously account for these data and
the high values of the large scale galaxy streaming
velocities suggested by the latest data.\ref{K.M. Gorski,
1992, Ap. J. Lett. 398, L5.}  However, only one channel of
the South Pole data were analyzed, with the signal in the
other three channels is interpreted as being galactic in
origin.\refmark{\rGaier}  It will be interesting to see
whether independent data sets show $\sim 1^\circ$ CMB
fluctuations at the level predicted by C+HDM.

Of the non-Gaussian models that have been proposed,\ref{See
e.g. L. Kofman \etal 1991, in {\it Proceedings, Workshop on
Large Scale Structure and Peculiar Motions in the Universe},
D.W. Latham and L.N. da Costa, eds. (Astronomical Society of
the Pacific), p. 251; D.S. Salopek 1992, Phys. Rev. D45,
1139.}  the idea of structure formation by
wakes of long cosmic strings is now perhaps the most
interesting one.  Cosmic strings and cosmic texture are both
generic, in the sense that particle physics Lagrangians with
suitable sets of scalar fields will automatically generate
such topological structures in the early universe.  The
texture model now appears to be ruled out by the COBE
data.\ref{D. Spergel, personal communication.} And the
version of cosmic strings that was most thoroughly
investigated, in which structure is seeded by small loops of
cosmic string, has now been ruled out since high resolution
simulations show that these loops do not survive long
enough: they are quickly cut up by string crossing and
reconnection.

The $\Omega=1$ long-string-wake Strings + Hot Dark Matter
model is well motivated and well specified---in
fact, it has only one parameter, the mass per unit length on the
string.  The dark matter in this scheme is presumably hot dark
matter: a $\tau$ neutrino with mass $m(\nu_\tau)$ given by
Eq. \omeganueq.  This (like the $m(\nu_\tau)$ needed for
C+HDM) is in the range suggested by the MSW explanation of
the solar neutrino data plus simple seesaw neutrino mass
models. With long string wakes providing the seeds for
structure formation, using hot rather than cold dark matter
gives this model relatively more large scale power and
is expected to suppress the formation of dense cores of dark
matter in galaxies. Preliminary investigation of this
scenario suggests that it might be consistent with COBE and
the large scale structure data (Bouchet, these proceedings).
More detailed calculations will be required to see whether
this is really true, and also whether the galaxies formed in
this model have the right properties and distribution.

My sketched Figure summarizes this discussion.  The three
most popular models for large scale structure of the
early-to-mid-1980s---HDM, CDM, and Cosmic String
Loops---are now all dead and buried (at least in their
simplest incarnations).  Let them rest in peace!  But from
their graves the three leading present models are growing:
CDM in an $\Omega\approx0.2$ universe with a cosmological
constant, $\Omega=1$ C+HDM, and $\Omega=1$ String Wakes
(with hot dark matter).  The former two are Gaussian models
consistent with cosmic inflation, the latter is a
non-Gaussian model that may\ref{H. Hodges and J.R. Primack
1991, Phys. Rev. D43, 3155.} be consistent with inflation.
If measurements of the cosmological paramenters turn out to
give low $\Omega$ and high $H_0=80-100$ km s$^{-1}$
Mpc$^{-1}$, then the first of these models is favored.  If
the indications from galaxy peculiar velocities and other
data that $\Omega\approx 1$ are valid, then the latter two
models are favored.  Of course, many other models have been
proposed, and many more are possible.  We have been
surprised before by the data and are likely to be surprised
again!

\noindent {\bf IV. Solar Neutrinos, Neutrino Masses, and Hot
Dark Matter}

The GALLEX intermediate-energy solar neutrino flux of
$83\pm19\pm8$ SNU (d'Angelo, these proceedings) is only a
little less than expectated in the Standard Solar Model
(SSM); and the entire SAGE dataset (Gavrin, these
proceedings) is not in disagreement with this. However, the
high-energy solar neutrino data from Kamiokande-III and
Homestake (Totsuko, Lande, these proceedings) are not
compatible with the SSM (Bahcall, Turck-Chi\`eze, these
proceedings).  The MSW neutrino-oscillation idea now seems
very attractive---and certainly less ad hoc than other
proposed solutions to the solar neutrino puzzle.  It is
interesting that it may also help explain why supernovae
explode\ref{G.M. Fuller \etal 1992, Ap. J. 389, 517} (see
Raffelt, these proceedings).

The MSW scheme requires that both the electron neutrino
$\nu_e$ and at least one other neutrino---say, the muon
neutrino $\nu_\mu$---have a nonvanishing mass, with
$m(\nu_\mu)>m(\nu_e)$.  Then the electron neutrinos emitted
in the center of the sun get an effective mass $m_{\rm
eff}(\nu_e)$ because of the high electron density there.
MSW also requires that $m(\nu_\mu)<m_{\rm eff}(\nu_e)$, and
that there be a nonvanishing mixing between $\nu_\mu$ and
$\nu_e$, analogous to the Cabibbo mixing between the first
two quark generations.  Then as the $\nu_e$'s stream out of
the center of the sun, $m_{\rm eff}(\nu_e)$ decreases and
eventually crosses $m(\nu_\mu)$.  As usual in
quantum-mechanical level crossing, the probability of
conversion of $\nu_e$ into $\nu_\mu$ will depend on the
$\nu_e$ energy and on the neutrino mixing and
Masses---actually on $m(\nu_\mu)^2-m(\nu_e)^2$. If we assume
that $m(\nu_\mu)>>m(\nu_e)$, then the combined solar
neutrino data imply that $$m(\nu_\mu) \approx (2-3)\times
10^{-3}\, {\rm eV}, \eqn\solarnu $$ with the $\nu_e \nu_\mu$
mixing angle $\theta_{e \mu}$ confined to two small regions,
either large or small (nonadiabatic) mixing ($\sin
2\theta_{e \mu} \approx 0.8$ or 0.1).

A muon neutrino mass in this range was expected in the
context of the ``seesaw'' mechanism for generating neutrino
masses,\ref{T. Yanagida 1978, Prog. Theor. Phys. B135, 66;
M. Gell-Mann, P. Raymond \& R Slansky 1979, in {\it
Supergravity} ed. P. van Nieuwenhuizen and D. Freedman
(North-Holland, Amsterdam), p. 315.} in which the light
left-handed neutrinos mix with heavy (mass $M$) right-handed
Majorana neutrinos.  The resulting neutrino masses are
related to the squares of the masses of the upper component
quarks of the same generations: $m(\nu_{e,\mu,\tau}) \approx
m^2_{u,c,t}/M$,\Ref\rNeuMod{For more detailed models, see
e.g. S.A. Bludman, D.C. Kennedy, and P.G. Langacker 1992,
Phys. Rev. D45, 1810; J. Ellis, J.L. Lopez \& D.V
Nanopoulos, 1992, Preprint CERN-TH.6569/92; and S.
Dimopoulos, L. Hall \& S. Raby 1992, preprint LBL-32484.}
so $$m(\nu_\tau)=\eta (m_t/m_c)^2 m(\nu_\mu),
\eqn\seesawnu$$ where $\eta\sim0.3$ is a model-dependent
factor including the effects of the running of coupling
constants. With $m(\nu_\mu)$ of Eq. (2) from the solar
neutrino data, and a top quark mass $\sim 10^2$ times that
of the charmed quark, this leads to $m(\nu_\tau) \sim 10$
eV, and correspondingly to a cosmological density of $\tau$
neutrinos again given by Eq. (1).  Even if the seesaw idea
is right, however, it remains an assumption of simplicity
that the heavy right-handed neutrinos in all three
generations have essentially the same mass $M$; if this
is not true, then the mass estimate for $m(\nu_\tau)$ above
is invalid.

Most exciting, the Chorus and Nomad $\nu_\mu \nu_\tau$
oscillation experiments now underway at CERN should see a
signal within about two years if these neutrino mass and
mixing models are right (Vanucci, these proceedings and \S V
below).

To summarize: the very plausible MSW explanation of the
solar neutrino data requires neutrino mass, and thereby goes
beyond the Standard Model of particle physics.  And the
combination of that, together with the admittedly rather
speculative seesaw model of neutrino masses with a single
intermediate mass scale $M$,  leads to the prediction that
light $\tau$ neutrinos---hot dark matter---may be all or at
least a considerable fraction of the dark matter.

\noindent {\bf V. Dark Matter Detection}

\noindent{A. Light neutrinos---new neutrino oscillation
experiments}

These speculations about neutrino masses can be tested by
experiments now underway at CERN, and also by using the next
generation of solar neutrino detectors, including
Super-Kamiokande and Sudbury Neutrino Observatory, to
clarify the energy-dependence and generational composition
of the solar neutrinos reaching the earth. By measuring the
energy spectrum of high energy solar neutrinos using both
charged and neutral current interactions, these experiments
can help determine whether the MSW model can explain the
data, and if so for which values of muon neutrino mass and
$\nu_e \nu_\mu$ mixing angle (Totsuka, Bahcall, these
proceedings).

Because they are less well known than the solar neutrino
experiments, I will describe here in a little more detail
the new $\nu_\mu \nu_\tau$ oscillation experiments now being
built at CERN (Vanucci, these proceedings).  In the Chorus
experiment (CERN-WA-095, approved September 1991), a beam of
muon neutrinos produced by the proton beam of the CERN-SPS
accelerator is directed at a target consisting of nearly a
ton of nuclear emulsion stacks.  If $m(\nu_\tau)$ lies in
the cosmologically interesting range from a few eV to $\sim
10^2$ eV and the $\nu_\mu \nu_\tau$ mixing angle $\theta$
satisfies $\sin^2 2 \theta \gsim 3\times 10^{-4}$, then a
substantial number of $\nu_\mu$ will oscillate to
$\nu_\tau$ on their way to the detector.  The emulsion stack
will then capture the $\sim 1$ mm tracks of the relativistic
$\tau$ leptons produced by $\nu_\tau +$ nucleon $\rightarrow
\tau^- + X$, thereby providing the first direct evidence for
the $\tau$ lepton as well as a measurement of its mass and
of $\theta$. The hard part is finding the tracks!  They are
to be located by a combination of techniques, including
scintillating fiber trackers, a layer of emulsion that is
changed biweekly, and a calorimeter that tags the $\tau^-$
decay by its transverse momentum imbalance.  The
complementary Nomad experiment (CERN-WA-096, also approved
September 1991) looks for $\tau$ production from the
$\nu_\mu$ beam in a 3 ton drift chamber target by a magnetic
detector based on the old CERN UA1 magnet. It searches for
the various decay modes of the $\tau$ using kinematical
criteria, and has roughly the same sensitivity as Chorus.

Are these experiments sensitive enough?  Yes, if the
$\nu_\mu \nu_\tau$ mixing angle $\theta$ is comparable to
the corresponding quark mixing angle
$\theta_{23}\approx0.04\pm0.01$, which would imply $\sin^2
2\theta \gsim$ few $\times 10^{-3}$.  Several particle
physics models of neutrino masses and mixings also suggest
that Chorus and Nomad may be sensitive
enough.\refmark{\rNeuMod} But even if $\theta$ is too small
for $\tau$s from $\nu_\mu \nu_\tau$ oscillations to be seen
in these CERN experiments, all is not lost: an emulsion
experiment similar to Chorus has been proposed at Fermilab
that will be an order of magnitude more sensitive after the
injector upgrade has been completed.  Thus it seems quite
likely that if $m(\nu_\tau)$ is in the range required for
Cold + Hot Dark Matter or Strings + Hot Dark Matter models,
this will soon be confirmed by accelerator experiments.

\noindent{B. WIMPs and Axions}

The cold dark matter particle candidates that are
well-motivated (in the sense that they have been proposed
for good particle-physics reasons independent of their
possible cosmological properties) are axions, still the best
solution to the strong-CP problem, and weakly interacting
massive particles (WIMPs), in particular the lightest
supersymmetric partner particle (LSP). Both are detectable
in the laboratory\ref{J.R. Primack, D. Seckel \& B.
Sadoulet 1988, Ann. Rev. Nucl. Part. Sci. 38, 751; P.F.
Smith and J.D. Lewin 1990, Phys. Rep. 187, 203.}  (Sadoulet,
these proceedings).

WIMPs that form the halo of our galaxy have an rms speed of
about $v=300$ km s$^{-1}=10^{-3} c$ (a little higher than
the 220 km s$^{-1}$ orbital speed of the sun in the
galactic disk, since the velocity of halo matter should be
essentially isotropic).  Thus a WIMP of mass $m$ has kinetic
energy $\half m v^2 = \half (mc^2/{\rm Gev})$ keV, which
can be transferred as recoil energy to a nucleus in a target
in the laboratory.  The probability of such interactions is
of course determined by the collision cross section;
non-detection of such events by ionization detectors has
already excluded large-cross-section WIMPs such as massive
Dirac neutrinos for a large range of masses.  A typical
event rate for allowed LSP WIMP dark matter particles is of
the order of an event per kg of target material per day, so
the problems are to have a large enough target to get a
decent event rate, and then to distinguish these dark matter
events from various backgrounds.  The group associated with
the Center for Particle Astrophysics at Berkeley has
demonstrated the efficacy of background rejection by
simultaneous detection of ionization and phonons from the
nuclear recoil in a germanium detector at a few millidegrees
K.  (Nuclear recoil produces a weaker ionization but a
stronger phonon signal, while the energy distribution is
just the opposite for the main background, electron recoil
from Compton scattering.) Other groups are pursuing other
approaches.  It is clear that detection of WIMP dark matter
will be hard, but it does appear to be technologically
feasible.  In a few years, experiments will either have
discovered the dark matter WIMP or ruled out a significant
part of the possible parameter space.

Axion searches attempt to detect the conversion of axions,
expected to have rest mass about $10^{-5}$ eV, into photons
of the same energy in a strong magnetic field.  (Despite
their low mass, axions are cold rather than hot dark matter
since they form as a non-thermal vacuum condensate.)  Such
searches have been conducted both at Brookhaven National
Laboratory and at the University of Florida, but their
sensitivity was about two orders of magnitude too low to
detect the predicted axion density. The detection
probability goes as the volume times the square of the
magnetic field, so it is hoped that the availability of a
large powerful magnet at Livermore National Laboratory will
permit a search with sufficient sensitivity to detect axion
dark matter or rule out essentially the entire expected mass
range.

\noindent{C. Searching for MACHOs by microlensing}

It is possibile that at least a fraction of the dark matter
in galaxy halos consists of some sort of astrophysical
objects. One possibility is that most of the ordinary matter
in the Universe may have been processed through a first
generation of ``Population III'' stars.  Dark remnants of
these objects have been termed ``Massive Astrophysical
Compact Halo Objects'' (MACHOs) by Kim Griest.

What form might the MACHOs take?  A variety of constraints
suggests that they might either be black hole remnants of a
population of ``Very Massive Objects'' (VMOs) larger than
100 $M_\odot$ that collapse entirely without ejecta during
their oxygen burning stage, or else objects that are too
small to burn hydrogen at all (brown dwarfs or jupiters). It
is also possible that they are primordial black holes.

If the MACHOs are VMO remnants, they would have two
important observational signatures: a background of
gravitational radiation generated by the black hole
collapses and a background of electromagnetic radiation
generated by the VMOs' nuclear-burning phase.  The
gravitational radiation may be detectable when laser
interferometers go on the air in the mid-1990s (see \S VI
below), but this depends on the very uncertain efficiency
with which collapsing VMOs produce gravitational waves.  The
electromagnetic radiation background might peak at 10
microns, where it would be hidden by zodiacal light; or it
could have been reprocessed by dust scattering, in which
case it would produce near-infrared spectral distortions
that the DIRBE instrument on COBE could observe,
anticorrelated in angle with submillimeter CMB
fluctuations.\ref{B.J. Carr \& J.R. Primack 1990, Nature,
345, 478; J.R. Bond, B.J. Carr \& C.J. Hogan 1991, Ap. J.
367, 420.}

The most interesting observational consequence of jupiters
would be their gravitational microlensing effects.  Lensing
occurs because light is bent in a gravitational field.
There are two distinct effects.  ``Macrolensing'' occurs
when the light from a distant object like a quasar is bent
by the gravity of an intervening galaxy or cluster to
produce multiple images. ``Microlensing'' occurs at similar
cosmological distances when an individual halo object
traverses one of the macrolensed images, thereby causing its
brightness to vary relative to the other images.  This is
only detectable at cosmological distances for halo objects
in the mass range below 0.1 $M_\odot$ because the timescale
of the fluctuation increases as the square root of the
deflector mass and exceeds an astronomer's lifetime above
0.1 $M_\odot$.  One can therefore use this effect to look
for jupiters but not VMO remnants. In fact, there is already
one claimed case of microlensing of a quasar,\ref{M.J. Irwin
\etal 1989, Astr. J. 98, 1989.} with the
deflector mass in the range 0.001 - 0.1 $M_\odot$.  However,
in this case the optical depth for microlensing was probably
greater than unity along the light path through the center
of the lensing galaxy, which makes the deduction of the
deflector mass uncertain.  Also, near a galaxy center the
most likely deflectors are ordinary matter rather than
MACHOs.

The most promising approach appears to be to seek
microlensing by MACHOs in our own Galactic halo by looking
for intensity variations of stars in the Large Magellanic
Cloud (LMC).\ref{B. Paczynski 1986, Ap. J. 304, 1; K. Griest
1991, Ap. J. 366, 412; K. Griest \etal 1991, Ap. j. Lett.
372, L79.}  In this case the
timescale of the variation is shorter, about a week for a
lensing object of 0.1 $M_\odot$.  It again varies as the
square root of the microlensing mass, so that it would be
practical to look for black hole remnants from VMOs as well
as jupiters.  However, the probability of a particular star
being microlensed is only $\sim10^{-6}$, so one therefore
has to look repeatedly at many stars.

Two groups have initiated searches for local microlensing.
A group of French astronomers and particle physicists,
represented at Blois by Ansari, has analysed Schmidt
telescope plates of the LMC, several hundred of which are
presently available, half taken since supernova 1987A.  Both
they and the Berkeley-Livermore-Mt. Stromlo (Australia)
collaboration are also repeatedly imaging the LMC with CCD
cameras on dedicated telescopes in the Southern
Hemisphere.  Since microlensing events with light
amplification factor $A$ are distributed uniformly in
$A^{-1}$ (which is proportional to the distance of closest
approach of the lensing object to the line of sight to the
lensed star), the large-$A$ events that will provide the
most convincing signal should not be that uncommon.  But
large-$A$ events have short duration, and the requirement of
frequent sampling favors the CCD approach over plates in
searching for jupiters. The fact that the increase in a
star's brightness due to microlensing is independent of
wavelength should help to distinguish lensing events from
intrinsic stellar variations, in which the color usually
changes with brightness.

\noindent {\bf VI. Getting Close to the Monsters}

By ``monsters'' I mean the compact objects such as black
holes that presumably power the astrophysical sources of
many of the highest energy particles.  Of course, the only
particles whose sources we can trace are those that travel
in straight lines---photons or neutrinos, or perhaps the
very highest energy charged particles.  The directions of
charged particles of lower energy are randomized by the
magnetic fields in our galaxy.  Since several detectors have
now seen high energy photons from the Crab pulsar, including
observations of TeV photons by the Whipple \v Cerenkov
telescope (Weekes, these proceedings), subsequently
confirmed by two French experiments, we know that neutron
stars can produce TeV particles.  (\v Cerenkov telescopes
observe the \v Cerenkov radiation produced in the atmosphere
by the particle showers initiated by energetic cosmic rays.)

EGRET, which has the highest energy sensitivity of the four
photon detectors on the Compton Gamma Ray Observatory, has
detected GeV photons from the directions of several
extragalactic sources (Strong, these proceedings), including
the quasars 3C373 and 3C279.  All these sources are Active
Galactic Nuclei (AGNs) of the ``blazar'' variety, thought to
be cases where the axis of powerful beamed radiation is
oriented almost directly along our line of sight.\ref{C.D.
Dermer \& R. Schlickeriser 1992, Science 257, 1642.}

One of the most exciting announcements at Blois92 was the
first detection of TeV extragalactic $\gamma$ rays, again by
the Whipple \v Cerenkov telescope (Lamb, these
proceedings).\ref{M. Punch \etal 1992, Nature 358, 477.}
The flux detected above 0.5 TeV was 0.3 that of the Crab
Nebula.  What was perhaps most notable about this discovery
was the fact that the source was not 3C279 but rather the
galaxy Mkn 421, the nearest of the EGRET extragalactic
sources, at a distance of about 90 $h^{-1}$ Mpc ($z=0.031$).
When it was flaring last spring, 3C279 was brighter than Mkn
421 at the GeV energies to which EGRET is sensitive, even
though 3C279, at a redshift $z=0.538$, is much farther away.
But the Whipple telescope could not detect 3C279.  An
appealing explanation is that the TeV photons from the more
distant sources were absorbed by $\gamma+\gamma
\rightarrow e^+ + e^-$ scattering on redshifted
starlight.  For typical estimates of the density of such
starlight, the universe will be optically thick to TeV
photons for redshift $z\gsim0.2$ (Stecker, these
proceedings).

A TeV photon scattering on an eV starlight photon has a
center-of-mass energy of order an MeV, just enough to
produce a $e^+ e^-$ pair.  The $e^+ e^-$ pair production
cross section is largest just above threshold, so the
absorption should be greatest for TeV energies.   This
explanation of the nonobservation of the more distant
sources of GeV photons can be confirmed by observing
high-energy photon sources at intermediate distances and at
higher energies, using \v Cerenkov telescopes and/or
extensive air shower arrays (Ong, these proceedings).  If it
turns out to be right, it opens a new sort of astronomy,
since this indirect observation of the density of redshifted
starlight can probe the era of galaxy formation. (I know of
no other way to measure the density of eV photons in the
universe.  We can't do it nearby, since the Milky Way is
itself a copious source of near-infrared photons.)

Yet another new sort of astronomy is suggested by these
observations. It is plausible that all EGRET AGNs are
powerful sources of high energy neutrinos as well as
photons.  The new large neutrino telescopes such as DUMAND
(in the sea near Hawaii) and AMANDA (in the ice at the South
Pole) should be able to detect TeV neutrinos from the same
sources.  The fluxes, energy spectra, and time behavior of
these high energy neutral particle signals should tell us
much about the physics in the exotic regions close to the
monsters (Stecker, these proceedings).

The directions of the highest energy cosmic rays, those with
$E \gsim 4\times 10^{19}$ eV, are not isotropic; the favored
direction is in the general direction of the Virgo cluster,
the nearest large concentration of galaxies (Cesarsky,
Cronin, these proceedings), although it is not entirely
clear that this is statistically significant when
observational biases are taken into account. Determining the
direction of the highest energy cosmic rays could clarify
both their composition and sources. Remarkably, it appears
to be possible to build a detector for such cosmic rays with
the huge collection area of $\sim$5000 km$^2$ for as little
as \$50 million (Cronin, these proceedings).  Depending on
the flux extrapolation, such a detector could collect
$\sim$5000 events per year with $E>10^{19}$ eV (five times
the total seen so far) and $\sim$5 per year with $E>10^{20}$
eV.

Although $\gamma$ ray bursts have long been known (they were
first discovered in the 1960s by satellites designed to
detect nuclear explosions on or near the earth), the
accumulating statistics from the BATSE detector on the
Compton GRO satellite have shown that their distribution is
remarkably isotropic.  This makes it unlikely that their
sources are galactic (unless they are close to the solar
system) or even in nearby galaxies, since the stars in our
galaxy and the nearby galaxies are both quite anisotropically
distributed. If the bursts arise at cosmological distances,
an interesting possibility is that they come from coalescing
binary neutron stars (see Glashow, these proceedings). We know
from binary pulsar observations that such systems exist.  Of
the 20 binary pulsars in our galaxy for which accurate
Kepler parameters have so far been determined, five are
binaries in which both members are neutron stars (Taylor,
these proceedings).  Reasonable estimates suggest that there
should be plenty of such systems out to the horizon to
account for the several hundred $\gamma$ ray bursts observed
per year.  A challenge for all models of $\gamma$ ray bursts
is to understand why so much of the energy emerges as
$\gamma$ rays.  The time dependence of the bursts is also
quite interesting, with several categories of bursts having
features from milliseconds to several seconds.

The news about binary pulsar observations is that they have
recently allowed a test of strong-field gravity (Taylor,
these proceedings).  Neutron stars have a surface
gravitational potential $GM/c^2 R\approx0.2$, $10^5$ times
larger than that at the surface of the sun. Some time ago,
binary pulsar data was used to confirm that binary pulsars
emit gravitational radiation at the expected rate; this test
is now at the 0.5\% level.  The latest work\ref{J.H. Taylor
\etal 1992, Nature 355, 132.} has measured
velocity-dependent and nonlinear gravitational phenomena
independently of the effects of gravitational radiation.
General relativity passes this new test perfectly.  The
current upgrade of the Arecibo telescope will permit still
more sensitive tests.

One of the most important announcements at Blois92 was the
commitment of funds from the French government for the
French-Italian VIRGO interferometric gravity wave
observatory. U.S. government funding for the LIGO detector
has also been made available.  These detectors should begin
to have sufficient sensitivity to see gravity waves,
although this may have to wait until advanced detectors
replace the first generation devices currently planned for
these observatories (Thorne, these proceedings). Several of
the phenomena I discussed above are possible sources for
such gravity waves, including VMO collapse and the ``last
three minutes'' of coalescing binary neutron stars.

\noindent {\bf VII. Conclusion}

This is certainly a golden age for cosmology and particle
astrophysics!  We are blessed with wonderful astronomical
instruments and accelerator experiments that each year open
new windows through which we see things that clarify the
initial conditions, the composition, and the evolution of
the universe.  The fact that the $\nu_e \nu_\mu$ oscillation
experiments now starting at CERN may confirm the prediction
of Cold+Hot Dark Matter that $m(\nu_\tau)\sim 7 h^2_{50}$ eV,
or perhaps that of Strings+HDM that $m(\nu_\tau)\sim 23
h^2_{50}$ eV, is a perfect illustration of the
interconnections growing between cosmology and particle
physics.

I have read in the newspapers that some in France are
unhappy with the rather bloody words of the {\it
Marseilleise}.  I will conclude this written version as I
concluded my summary talk at Blois, with some new words to
the same stirring melody:

\bs
\centerline{\it ``Bloiseilleise''}
\bs

\line{{\hskip 2in} Cosmologists and astrophysicists, \hfil}
\line{{\hskip 2in} The day of glory will soon arrive! \hfil}
\line{{\hskip 2in} While we outline the tasks still before us, \hfil}
\line{{\hskip 2in} We have fed well but ignored our wives \hfil}
\line{{\hskip 2in} We have fed well but ignored our wives. \hfil}
\smallskip
\line{{\hskip 2in} They may think we have no sensitivity... \hfil}
\line{{\hskip 2in} But in detection of dark matter, yes we do! \hfil}
\line{{\hskip 2in} We'll find out what the universe is made of, \hfil}
\line{{\hskip 2in} And eliminate those factors of two. \hfil}
\smallskip
\line{{\hskip 2in} To your pencils, colleagues at Blois! \hfil}
\line{{\hskip 2in} Frame a theory elegant and clear! \hfil}
\line{{\hskip 2in} March on / March on \hfil}
\line{{\hskip 2in} We'll diet when we're gone, \hfil}
\line{{\hskip 2in} And give our thanks to Tran! \hfil}

\bs\bs
\noindent {\bf Acknowledgments}

To Jean Tran Thanh Van for the excellent arrangements
at Blois, and to the participants for their patience when I
pestered them for further explanations---although they are of
course not to blame for whatever I still didn't understand
as I prepared this summary! I thank my wife Nancy Abrams for
her help with the {\it ``Bloiseilleise.''} I appreciate the
travel funds and warm hospitality I received from Saclay
and from the Institut d'Astrophysique, Paris, where I wrote
the first draft of this written version. I also acknowledge
support from NSF and faculty research grants at UCSC.

\bs\bs
\noindent {Caption for Cartoon Figure---}
Standard CDM with constant linear bias $b$ is inconsistent
with the data, as is standard HDM. The idea that loops of
cosmic string seed structure formation has been killed by
high-resolution simulations showing that the loops do not
survive long enough.  But from the graves of these models
potentially successful models are growing: CDM with
$\Omega\approx0.2$ and a cosmological constant, and
$\Omega=1$ Cold+Hot Dark Matter and Strings + Hot Dark
Matter.

\endpage

\bs
\refout

\bye